\documentclass[prb,endfloats,eqsecnum]{revtex4}
\usepackage{graphicx}
\newcommand{\bl}{{\bf J}}
\newcommand{\bk}{{\bf K}}
\newcommand{\bm}{{\bf M}}

\newcommand{\B}{{\bf B}}
\newcommand{\E}{{\bf E}}
\newcommand{\F}{{\bf F}}
\newcommand{\hf}{\hat{\bf F}}
\newcommand{\he}{{\bf e}}
\newcommand{\hb}{{\bf b}}

\newcommand{\bh}{\hat{\bf B}}
\newcommand{\tm}{\xi}

\newcommand{\br}{{\bf r}}

\newcommand{\bu}{{\bf u}}

\newcommand{\ep}{\varepsilon}

\newcommand{\e}{{\rm e}}

\newcommand{\nn}{\nonumber}
\newcommand{\la}{\label} 

\newcommand{\be}{\begin{equation}}
\newcommand{\ee}{\end{equation}}
\newcommand{\ba}{\begin{eqnarray}}
\newcommand{\ea}{\end{eqnarray}}

\begin{document}
\title{Relativistic Motion in a Constant Electromagnetic Field}

\author{Siu A. Chin}

\affiliation{Department of Physics, Texas A\&M University,
College Station, TX 77843, USA}

\begin{abstract}
For a relativistic charged particle moving in a constant 
electromagnetic field, its velocity 4-vector has been well studied. However,
despite the fact that both the electromagnetic field and the equations
of motion are purely real, the resulting 4-velocity is 
seemingly due to a complex electromagnetic field.
This work shows that this is not due to some complex formalism used
(such as Clifford algebra) but is intrinsically due to the
fact that the $o(3,1)$ Lie algebra of the Lorentz group is 
equivalent to two commuting complex $su(2)$ algebras. Expressing
the complex $su(2)$ generators in terms of the boost and 
rotation operators then naturally introduces a complex electromagnetic 
field. This work solves the equation of motion not as a matrix equation, but
as an operator evolution equation in terms of the  
generators of the Lorentz group. The factorization of the
real evolution operator into two commuting complex evolution operators 
then directly gives the time evolution of the velocity 4-vector without 
any reference to an intermediate field.

\end{abstract}
\maketitle

\section {Introduction}

The equations of motion of a negatively charged particle
$q=-e$ moving in a constant electromagnetic field
$F_{\mu\nu}$ is given by
\be
\frac{du_\mu}{d\tau}=\frac{e}{mc} u_\nu F_{\nu\mu}
\la{vel}
\ee
and 
\be
\frac{dx_\mu}{d\tau}=u_\mu
\la{pos}
\ee
where $x_\mu=(x_0,\br)$, $x_0=ct$, $u_\mu=(u_0,\bu)$ and $\tau$ is the
proper time. Once (\ref{vel}) is known, the integration of (\ref{pos})
is straightforward. This is a well studied problem, with many
published solutions\cite{taub48,hell55,hy97,bacry70,hest74,sal85,bay89,zen90,zen92}. 
Since (\ref{vel}) is just a matrix equation,
it can be directly integrated, as done by Taub\cite{taub48}, Hellwig\cite{hell55}
and Hyman\cite{hy97}. However, the algebraic manipulations used were purely  
formal and gave no insight on why the solution is the way it is.
Bacry, Combe and Richard\cite{bacry70} have offered a group theoretic 
analysis of the problem, but the explicit form of 
the 4-velocity solution was not given. More recently, Hestenes\cite{hest74}, 
Salingaros\cite{sal85}, Baylis and Jones\cite{bay89} and Zeni and Rodrigues\cite{zen90,zen92}
have applied Clifford algebra techniques to solve this problem.
Salingaros has given a detailed comparison with earlier results of Taub and Hestenes,
correcting some discrepancies.	However, his explicit solution is only for a particle 
starting from rest. Baylis and Jones' solution for an arbitrary initial 4-velocity is  
very abbreviated. Zeni and Rodrigues' solution is similar in form to the solution 
presented here, but the derivations are completely different.

Salingaros' work highlighted the fact that, despite the equation of 
motion (\ref{vel}) and the electromagnetic field are both real, 
the resulting 4-velocity is seemingly due to a complex electromagnetic field. 
Since complex fields are a natural part of Clifford algebra\cite{hest74,sal85,bay89,zen90,zen92}, 
it is unclear whether the solution merely reflect the formalism used or that 
a complex electromagnetic field is an intrinsic part of the solution.

In this work, we solve (\ref{vel}) by an entirely elementary method without
invoking any advance formalism such as Clifford algebra. Instead
of solving (\ref{vel}) as a matrix equation, we solve it via  
an evolution operator, as it is done in the Poisson bracket formulation
of symplectic integrators\cite{dragt,neri87,fr90,yos93,mcl95}. Just like the quantum evolution operator, 
which acts on any quantum state and evolve it forward in time, the classical 
evolution operator acts on any dynamincal variable and evolve it forward in time.
The method of classical evolution operator is described in Section II. 
For a relativistic particle in a constant electromagnetic field, the 
corresponding evolution operator is the exponential of 
generators of the Lorentz group.  Since the $o(3,1)$ Lie algebra of the Lorentz 
group generators is equivalent to two commuting complex $su(2)$ algebras, the
evolution operator can be factored into two complex evolution
operators. In each complex evolution operator, reexpressing
the complex $su(2)$ generators as boost and rotation operators then naturally introduces 
a complex electromagnetic field. Thus the complex electromagnetic field is
an intrinsic part of the solution due to the structure of the Lorentz group,
and is independent of any Clifford algebra formalism.
This is shown in Section III. The explicit velocity solution, for an arbitrary
initial 4-velocity vector, is given in Section IV. Some conclusions and applications
are indicated in Section V.

\section {Classical evolution operator}

Eq.(\ref{vel}) can be written out
explicitly as
\ba
\frac{d\bu}{d\tau}&=&\frac{e}{mc}(\B\times\bu -\E u_0)\la{uvec}\\ 
\frac{du_0}{d\tau}&=&\frac{e}{mc}( -\E\cdot \bu).
\la{uzero}
\ea
The evolution of any 
dynamical variable $W(x_\mu,u_\mu)$ (including $x_\mu$ and $u_\mu$ themselves), 
is given by
\ba
\frac{dW}{d\tau}&=&\frac{\partial W}{\partial x_\mu}\frac{dx_\mu}{d\tau}
+\frac{\partial W}{\partial u_\mu}\frac{du_\mu}{d\tau}\la{poisson}\\
&=&\Bigl(u_\mu\frac{\partial}{\partial x_\mu}+
\frac{e }{mc}u_\nu F_{\nu\mu}\frac{\partial}{\partial u_\mu}\Bigr)W=(T+V)W. 
\la{evol}
\ea
If $dx_\mu/d\tau$ and $du_\mu/{d\tau}$ are derivable from a Hamiltonian,
then the RHS of (\ref{poisson}) is just the Poisson bracket. However, as realized in
Ref.\cite{chin081}, as long as one has the equations of motion, one can define an 
evolution operator regardless whether one has a canonical Hamiltonian. 
Eq.(\ref{evol}) has the operator 
solution,
\be
W(x_\mu(\tau),u_\mu(\tau))=\e^{\tau(T+V)}W(x_\mu,u_\mu).
\la{genev}
\ee
In particular, because $F_{\mu\nu}$ is constant,
$(T+V)^nu_\mu=V^nu_\mu$, the 4-velocity is evolved simply by
\be
u_\mu(\tau)=\e^{\tau(T+V)}u_\mu=\e^{\tau V}u_\mu.
\la{vev}
\ee
From the explicit forms (\ref{uvec}) and (\ref{uzero}), one sees that
\ba
\tau V&=&\frac{e\tau }{mc}\Bigl(
(\B\times\bu)\cdot\frac\partial{\partial \bu}
-\E\cdot \bu\frac\partial{\partial u_0}
-u_0\E \cdot\frac\partial{\partial \bu}
\Bigr)\nn\\
&=&\tm(\B\cdot\bl+\E\cdot\bk)
\ea
where $\tm=e\tau/mc$ is the scaled proper time, and
\be
\bl=\bu\times \frac{\partial}{\partial \bu}
\qquad
\bk=-\bu\frac{\partial}{\partial u_0}-u_0\frac{\partial}{\partial \bu}
\la{lgen}
\ee
are the three rotation and three boost generators of the Lorentz group.
If $\E=0$, then
\be
u_0(\tm)=\exp\left(\tm\B\cdot\bl\right)u_0=u_0
\la{vb0}
\ee
\ba
\bu(\tm)&=&\exp\left(\tm\B\cdot\bl\right)\bu\nn\\
&=&\exp\left(\tm B(\bh\times\bu)\cdot\frac{\partial}{\partial \bu}\right)\bu
\la{bform}
\ea
where $\hat\B=\B/B$ is the unit vector of the magnetic field.
Using the above form (\ref{bform}), as shown in Ref.{\cite{chin081}, 
the expansion of the exponential 
then directly gives the finite rotation
\be
\bu(\tm)
=\bu+\sin(\tm B)(\bh\times\bu) +
(1-\cos(\tm B))\bh\times(\bh\times\bu).
\la{vb}
\ee
Similarly, if $\B=0$, then the
exponential operator gives 
\ba
u_0(\tm)&=&\exp\left(\tm\E\cdot\bk\right)u_0\nn\\
&=&\Bigl(1+\frac12 (\tm E)^2+\cdots\Bigr)u_0
-\Bigl(\tm E+\frac1{3!}(\tm E)^3+\cdots\Bigr)\hat\E\cdot\bu\nn\\
&=&\cosh(\tm E)u_0-\sinh(\tm E)\hat\E\cdot\bu\la{ve0}\\
\bu(\tm)&=&\exp\left(\tm\E\cdot\bk\right)\bu\nn\\
&=&\bu+
\hat\E\Bigl((\cosh(\tm E)-1)\hat\E\cdot\bu-\sinh(\tm E)u_0\Bigr).
\la{ve}
\ea
Eqs.(\ref{ve0})-(\ref{ve}) correspond to a general Lorentz boost in the direction 
of $\hat\E=\E/E$ preserving the Minkowski norm $u_0^2(\tm)-|\bu(\tm)|^2=u_0^2-|\bu|^2$. 
The reason for the choice of $q=-e$ is that the rotation is then right-handed and 
the Lorentz boost is the standard transformation rather than its inverse.

If both $\B$ and $\E$ are non-vanishing, then the general solution for the
velocity 4-vector is given by
\be
 u_\mu(\tm)=\e^{\tm(\B\cdot\bl+\E\cdot\bk)}u_\mu.
 \la{evoper}
\ee
From the study of symplectic integrators\cite{dragt,neri87,fr90,yos93,mcl95},
such an exponential of two operators can be approximated to any order in $\tm$ via
the product decomposition
\be
\e^{\tm(\B\cdot\bl+\E\cdot\bk)}=\prod_i\e^{b_i\tm\B\cdot\bl}\e^{e_i\tm\E\cdot\bk}
\la{prod}
\ee
with suitable coefficients $\{b_i,e_i\}$.
Since the effect of each exponential is known (\ref{vb0})-(\ref{ve}), the the general 
product can be computed in sequence.
The two elementary second order approximations are:
\ba
\e^{\tm(\B\cdot\bl+\E\cdot\bk)}
&=&\e^{(\tm/2)\B\cdot\bl}\e^{\tm\E\cdot\bk}\e^{(\tm/2)\B\cdot\bl}+O(\tm^3)\nn\\
&=&	\e^{(\tm/2)\E\cdot\bk}\e^{\tm\B\cdot\bl}\e^{(\tm/2)\E\cdot\bk}+O(\tm^3).
\ea
This approach is purely real, no complex quantity enters anywhere. However,
since we are interested only in the exact evaluation of (\ref{evoper}),
higher order approximations of the form (\ref{prod}) will not be considered here. 

\section {Complex decomposition}

The Lorentz group generators (\ref{lgen}) obey the well known commutator
relations:
\be
[J_i,J_j]=-\ep_{ijk}J_k \qquad[J_i,K_j]=-\ep_{ijk}K_k \qquad[K_i,K_j]=\ep_{ijk}J_k.
\ee
This implies that
\be
[\B\cdot\bl,\E\cdot\bk]=(-\B\times\E)\cdot\bk.
\ee
Thus if $\B\times\E=0$, then $\B\cdot\bl$ and $\E\cdot\bk$ commute and 
the solution is given
by
\be
 u_\mu(\tm)=\e^{\tm\B\cdot\bl}\e^{\tm\E\cdot\bk}u_\mu
\la{para}
\ee
This is the common starting point of many solutions\cite{hest74,sal85}. 
The remaining task is then to transform to such a Lorentz frame in which 
$\B\times\E=0$, apply the solution (\ref{para})
and transform back\cite{hest74,sal85}. However, there is a much more direct way of decomposing 
$\B\cdot\bl+\E\cdot\bk$
into two commuting operators based on the fact that
the Lie algebra of $o(3,1)$ is equivalent to that of $su(2)\times su(2)$.
The complex operators
\be
\bm=\frac12(\bl+i\bk) \qquad {\rm and}\qquad \bm^*=\frac12(\bl-i\bk)
\ee     
have commutators,
\be
[M_i,M_j]=-\ep_{ijk}M_k \qquad[M^*_i,M^*_j]=-\ep_{ijk}M^*_k \qquad[M_i,M^*_j]=0.
\ee
Thus we have an exact, but complex decomposition 
\be
\e^{\tm(\B\cdot\bl+\E\cdot\bk)}= \e^{\tm(\B+i\E)\cdot\bm^*}\e^{\tm(\B-i\E)\cdot\bm}.
\la{complex}
\ee
In (\ref{complex}), one notices that
\be
\e^{\tm(\B-i\E)\cdot\bm}=\e^{\tm(\B-i\E)\cdot(\bl+i\bk)/2}
\ee
corresponding to having a complex electric field $\F/2=(\E+i\B)/2$
and a complex magnetic field $-i\F/2=(\B-i\E)/2$. Since
$-i\F\times \F=0$, one can further decompose, 
\be
\e^{\tm(\B-i\E)\cdot(\bl+i\bk)/2}
=\e^{\tm (\F/2)\cdot\bk}\e^{\tm(-i\F/2)\cdot\bl}.
\la{msplit}
\ee
Similarly, 
\be
\e^{\tm(\B+i\E)\cdot(\bl-i\bk)/2}
=\e^{\tm(-i\F/2)^*\cdot\bl}\e^{\tm(\F/2)^*\cdot\bk}.
\la{mstar}
\ee
Thus the complex decomposition of the purely real
evolution operator $\exp[\tm(\B\cdot\bl+\E\cdot\bk)]$
naturally introduces the complex electromagnetic field
$\F=\E+i\B$. Furthermore, the evolution operator consists of
four pieces, best evaluated in pair as in (\ref{msplit}) and
(\ref{mstar}). Note that
\be
\e^{\tm(\F/2)^*\cdot\bk}\e^{\tm (\F/2)\cdot\bk}\neq	\e^{\tm \E\cdot\bk}
\ee 
and the evaluation of (\ref{msplit}) and (\ref{mstar}) cannot be
further simplified. The complex decomposition has been used by Fredsted\cite{fred01}
to obtain a spinor representation of the Lorentz group, but he
did not solve the charged particle problem as it is done here.

\section {Explicit solution}

In order to evaluate (\ref{msplit}) and (\ref{mstar}) using
our previous results (\ref{vb0})-(\ref{ve}), it is only necessary to define
the norm of the complex vector $\F=\E+i\B$ as
\be
F=\sqrt{\F\cdot\F}=\sqrt{|\E|^2-|\B|^2+i 2(\E\cdot\B)}
\ee
and the corresponding complex unit vector as $\hat\F=\F/F$.
For $z=x+iy=|z|e^{i\theta}$, $\sqrt{z}=u+iv$, where
\ba
u&=&|z|^{1/2}\cos(\theta/2)=\sqrt{|z|(1+\cos\theta)/2}=\sqrt{(|z|+x)/2}\nn\\
v&=&|z|^{1/2}\sin(\theta/2)=\sqrt{|z|(1-\cos\theta)/2}=\sqrt{(|z|-x)/2}.
\ea
For the ease of later comparison, we will follow Salingaros' notation and define
\be
\kappa_1=|\E|^2-|\B|^2\qquad \kappa_2=2(\E\cdot\B)\qquad\kappa=\sqrt{\kappa_1^2+\kappa_2^2}
\ee
\be
E^\prime=\sqrt{\frac{\kappa+\kappa_1}2}\quad{\rm and}\quad B^\prime=\sqrt{\frac{\kappa-\kappa_1}2}
\ee
Thus
\be 
F=E^\prime+iB^\prime
\quad{\rm and}\quad F^*F=(E^\prime)^2+(B^\prime)^2=\kappa
\ee
Since two exponential operators in (\ref{msplit}) commutes,
their effects simply add:
 \begin{equation}
 \left(\begin{array}{c}
       u_0^\prime\\
       \bu^\prime
      \end{array}\right)
=
\e^{\tm (\F/2)\cdot\bk}\e^{\tm(-i\F/2)\cdot\bl}
\left(\begin{array}{c}
       u_0 \\
       \bu
      \end{array}\right)
\label{fhalf}
\end{equation}
resulting in
\ba
u_0^\prime&=&\cosh(\tm F/2)u_0-\sinh(\tm F/2)\hat\F\cdot\bu\nn\\
\bu^\prime&=&\bu
+\hat\F\Bigl((\cosh(\tm F/2)-1)\hat\F\cdot\bu-\sinh(\tm F/2)u_0\Bigr)\nn\\
&&+\sin(-i\tm F/2)(\hat\F\times\bu) +
(1-\cos(-i\tm F/2))\hat\F\times(\hat\F\times\bu)
\ea
The latter now {\it greatly} simplifies to
\be
\bu^\prime
=\cosh(\tm F/2)\bu-\sinh(\tm F/2)(u_0\hat\F+i\hat\F\times\bu).
\ee
One can check that the cross-product term above is essential for
preserving the Minkowski norm $(u_0^\prime)^2-(\bu^\prime)^2=u_0^2-\bu^2$. 
To further simplify the notation, let's denote
\be
c=\cosh(\tm F/2)\quad{\rm and}\quad s=\sinh(\tm F/2) 
\ee
then, finally,
\begin{equation}
\left(\begin{array}{c}
       u_0(\tm) \\
       \bu(\tm)
      \end{array}\right)
=\e^{\tm(-i\F/2)^*\cdot\bl}\e^{\tm(\F/2)^*\cdot\bk}.
\left(\begin{array}{c}
       u_0^\prime\\
       \bu^\prime
      \end{array}\right)
\label{shalf}
\end{equation}
where
\ba						  
u_0(\tm)&=&c^*u_0^\prime-s^*\hat\F^*\cdot\bu^\prime\nn\\
 &=& (c^*c+s^*s\hf^*\cdot\hf)u_0-\Bigl(2{\rm Re}(c^*s\hf)-i s^*s\hf^*\times\hf\Bigr)\cdot\bu\\
\bu(\tm)&=&c^*\bu^\prime
-s^*(u_0^\prime\hat\F^*-i\hat\F^*\times\bu^\prime)\nn\\
&=&	(c^*c-s^*s\hf^*\cdot\hf)\bu-\Bigl(2{\rm Re}(c^*s\hf)+i s^*s\hf^*\times\hf\Bigr)u_0\nn\\
  &&+2{\rm Im}(c^*s\hf)\times\bu+s^*s(\hf^*\hf+\hf\hf^*)\cdot\bu.
\ea
All terms can now be easily computed:
$$
c^*c=\frac{\cosh(\tm E^\prime)}2+\frac{\cos(\tm B^\prime)}2 \qquad
s^*s=\frac{\cosh(\tm E^\prime)}2-\frac{\cos(\tm B^\prime)}2 
$$
$$
c^*s=\frac{\sinh(\tm E^\prime)}2+i\frac{\sin(\tm B^\prime)}2 \qquad \hf=\he+i\hb
$$
$$
\he=\frac1\kappa(E^\prime\E+B^\prime\B) \qquad \hb=\frac1\kappa(E^\prime\B-B^\prime\E)
$$
$$
\hf^*\cdot\hf=\frac{E^2+B^2}\kappa \qquad   \hf^*\times\hf=2i\frac{\E\times\B}\kappa \qquad
 \hf^*\hf+\hf\hf^*=2\frac{\E\E+\B\B}\kappa
$$
$$
2{\rm Re}(c^*s\hf)
=\frac\E\kappa [E^\prime\sinh(\tm E^\prime)+B^\prime\sin(\tm B^\prime)]
+\frac\B\kappa [B^\prime\sinh(\tm E^\prime)-E^\prime\sin(\tm B^\prime)]
$$
$$
2{\rm Im}(c^*s\hf)
=\frac\E\kappa [E^\prime\sin(\tm B^\prime)-B^\prime\sinh(\tm E^\prime)]
+\frac\B\kappa [B^\prime\sin(\tm B^\prime)+E^\prime\sinh(\tm E^\prime)].
$$
Putting everything together then yields
\ba						  
u_0(\tm)&=&\Bigl(\frac{\cosh(\tm E^\prime)}2+\frac{\cos(\tm B^\prime)}2
+\frac{E^2+B^2}{\kappa}\Bigl[\frac{\cosh(\tm E^\prime)}2-\frac{\cos(\tm B^\prime)}2\Bigr] \Bigr)u_0\nn\\
 &&-\Bigl(\frac\E\kappa [E^\prime\sinh(\tm E^\prime)+B^\prime\sin(\tm B^\prime)]
+\frac\B\kappa [B^\prime\sinh(\tm E^\prime)-E^\prime\sin(\tm B^\prime)]\nn\\ 
 &&\qquad\qquad\qquad\qquad\qquad\qquad\qquad\quad+\frac{\E\times\B}\kappa [\cosh(\tm E^\prime)
 -\cos(\tm B^\prime)]\Bigr)\cdot\bu\la{u0p}\\
\bu(\tm)&=& \Bigl(\frac{\cosh(\tm E^\prime)}2+\frac{\cos(\tm B^\prime)}2
-\frac{E^2+B^2}{\kappa}\Bigl[\frac{\cosh(\tm E^\prime)}2-\frac{\cos(\tm B^\prime)}2\Bigr] \Bigr)\bu\nn\\
&&-\Bigl(\frac\E\kappa [E^\prime\sinh(\tm E^\prime)+B^\prime\sin(\tm B^\prime)]
+\frac\B\kappa [B^\prime\sinh(\tm E^\prime)-E^\prime\sin(\tm B^\prime)]\nn\\ 
 &&\qquad\qquad\qquad\qquad\qquad\qquad\qquad\quad-\frac{\E\times\B}\kappa [\cosh(\tm E^\prime)
 -\cos(\tm B^\prime)]\Bigr)u_0                                        \nn\\
  &&+\Bigl(\frac\E\kappa [E^\prime\sin(\tm B^\prime)-B^\prime\sinh(\tm E^\prime)]
+\frac\B\kappa [B^\prime\sin(\tm B^\prime)+E^\prime\sinh(\tm E^\prime)]   \Bigr)\times\bu\nn\\
  &&\qquad\qquad+[\cosh(\tm E^\prime)-\cos(\tm B^\prime)]\frac{\E(\E\cdot\bu)+\B(\B\cdot\bu)}\kappa. 
\la{bup}   
\ea
The correctness of (\ref{u0p}) and (\ref{bup}) can be checked in four ways.
First and second, for $\E=0$ and $\B=0$, (\ref{u0p})-(\ref{bup}) reproduces 
(\ref{vb0})-(\ref{vb}) and (\ref{ve0})-(\ref{ve}) respectively. 
Third, for $\E\times\B=0$, (\ref{u0p}) and (\ref{bup}) are identical to
(\ref{para}). Fourth, if initially
$\bu=0$ and $u_0=1$ ($c=1$), then (\ref{u0p}) and (\ref{bup}) are in complete 
agreement with Salingaros's result\cite{sal85}, his Eq.(28). 
(Recalls that we take $q=-e$, $\tm=e\tau/m$ here is
the negative of Salingaros' $\zeta=q\tau/m$.) Thus every term in (\ref{u0p}) and (\ref{bup})
has been verified in at least one special case.

\section {Conclusions}

In this work, we have solved the problem of relativisitc motion			   
in a constant electromagnetic field by the method of classical
evolution operator. The distinct advantage of this approach is that
the resulting evolution operator is just
the {\it finite} group generator of the Lorentz group and the
fundamental group structure of the Lorentz group can be fully
exploited for its evaluation. The resulting complex decomposition
explains why the solution, despite being real, is fundamentally
due to a complex electromagnetic field. 

In contrast to other methods, which can only solve the problem
when the electromagnetical field is constant, 
our operator approach can be extended to solve the case when 
the electromagnetic field is no longer spatially uniform. 
In this general case, the full evolution operator (\ref{genev}) can 
still be approximated 
to any order via
\be
W(x_\mu(\tau),u_\mu(\tau))=\prod_{i}\e^{t_i\tau T}\e^{v_i\tau V}W(x_\mu,u_\mu),
\la{qvev}
\ee
for suitable sets of coefficients $\{t_i,v_i\}$. Since
the effect of both $\e^{\tau T}$ and $\e^{\tau V}$ are known,
any dynamical variable $W$ can be evolve forward in $\tau$ by 
sequentially updating $x_\mu$ and $u_\mu$ via
\be
u_\mu(v_i\tau)=\e^{v_i\tau V}u_\mu
\ee
and
\be
x_\mu(t_i\tau)=\e^{t_i\tau T}x_\mu=x_\mu+t_i\tau u_\mu.
\ee
The detail of solving relativistic motion in a non-uniform 
field will be given in a future study.


\newpage
\centerline{REFERENCES}

\end{document}